\documentclass[a4paper,8pt]{article}
\usepackage[cp1251]{inputenc}
\usepackage[russian,english]{babel}
\usepackage{array,longtable}
\usepackage{amssymb, latexsym, amsthm, amsmath, amsxtra}
\usepackage{graphicx}

\graphicspath{{}}
\DeclareGraphicsExtensions{.pdf,.png,.jpg,}

\makeatletter

\begin{document}
\large

\begin{center}
\title{}{\bf   About the corollaries of the soliton-breather models of particles tunneling on the example of the nonlinear Klein-Gordon and Schrodinger equation.

 }
\vskip 1cm

\author{}
R.K. Salimov \textsuperscript{1}, E.G. Ekomasov \textsuperscript{1}
{}

\vskip 1cm

{ \textsuperscript{1} Bashkir State University, 450076, Ufa, Russia }


\vskip 0.5cm
e-mail: salimovrk@bashedu.ru

\end{center}

\vskip 1cm

{\bf Abstract}
 \par
 The paper presents soliton-breather models of particles tunneling on the example of Klein-Gordon and Schrodinger equation nonlinear breathers. It is shown that in this case the non-linearity registration should lead to spatial restrictions in breathers wave properties observation and to appearance of radiation during tunneling. The paper also presents a new type of Schrodinger equation, which leads to the solutions localization in a space three-dimensional spherically symmetric case.

 \par
 \vskip 0.5cm


\par
\vskip 1cm

Nonlinear wave equations, such as Klein-Gordon nonlinear equations, are used in many areas of physics, including hydrodynamics, condensed matter physics, field theory, etc. [1-8]. The study of nonlinear scalar fields is also topical in cosmology [9-11]. Stable solutions of these equations can be interpreted as the classic models of the final size particles. For example, Sin-Gordon equation was used by Skyrme as a model in the nonlinear theory of field [12]. Skyrme suggested this equation as a nonlinear generalization of Klein-Gordon equation. In his theory solitons were interpreted as elementary particles, rather than a soliton part of solution, consisting of small-amplitude waves – as bosons, that mediate the interaction. Hereinafter solitons are understood as localized particle-like solutions that do not necessarily exhibit the soliton properties at collisions. When considering solitons as particles the biggest interest represents a spatially three-dimensional case. However, the generalization of Klein-Gordon equation stationary localized solitons to the three-dimensional case is prohibited by Derrick theorem [13]. Moreover, non-stationary solutions in the form of breathers also appear unstable for the case of sine-Gordon equation [14]. Stable breather or pulson solutions in a three-dimensional case belong to the equations that have no linear limit as solution amplitude tends to zero. An example of such equations can be equations of the form with the fractional degree nonlinearity:

\begin{align}
  u_{rr}+2 \frac{u_r}{r} -u_{tt}=u^{\frac{m}{n}}
   ,\label{eq:1}
   \end{align}

where $n>m;n=2k+1;k=1,2,3,...;m=2l+1;l=0,1,2,3,...$. These equations have stable solutions in the form of pulsons [15]. The stability of these solutions can be explained by the fact that at the conversion of the form   equation (1) is transformed into
\begin{align}
  v_{rr} -v_{tt}=r^{\frac{n-m}{n}}v^{\frac{m}{n}}    \label{eq:1}
   \end{align}

From the equation (3) one can see that the spreading of localized solutions is prevented by the field of form $r^{\frac{n-m}{n}}v^{\frac{m}{n}}$ . If the solution   does not spread, the solution   will not spread as well. But such approach makes impossible any radiation from a localized solution. In order to introduce in a similar model any radiation, we can consider a Lorentz-invariant model of the form
\begin{align}
u_{xx}+u_{yy}+u_{zz}-u_{tt}=F(u)G(u_{x}^2+u_{y}^2+u_{z}^2-u_{t}^2)
 ,\label{eq:1}
   \end{align}
In particular the equation
\begin{align}
u_{xx}+u_{yy}+u_{zz}-u_{tt}=u^{\frac{m}{n}}(u_{x}^2+u_{y}^2+u_{z}^2-u_{t}^2)^{\frac{k}{s}}
 ,\label{eq:1}
   \end{align}
  Where  $m,n,s$- odd numbers, $k$ - even number and  . Localized solutions of such model will be stable for a spherically symmetric case, as well as for model (3). For example, for the equation

\begin{align}
  u_{rr}+2 \frac{u_r}{r} -u_{tt}=u^{\frac{3}{13}}(u_r^2-u_t^2)^{\frac{3}{13}}
   ,\label{eq:1}
   \end{align}
it was shown numerically in paper [16]. Such equation will also have solutions in the form of plane waves and possibly their localized counterparts. Thus, in such model a localized pulson solution can be treated in the form of elementary particles and waves, for which the right-hand side of equation (4) becomes zero, may be considered as radiation. Such an interpretation is similar to Skyrme interpretation of the sin-Gordon equation solutions. Although breathers in a spatially one-dimensional case are homotopically indistinguishable from the vacuum solution, in a three-dimensional case they can possibly have a nontrivial topology, necessary for the description of charges conservation. This approach is interesting in that consideration of pulsons as particle provides in a natural way to get a spatial modulation of solutions, similar to de Broglie wave modulation. An example of this can be a sine-Gordon equation breather, which for zero speed is of the form
   \begin{align}
u=4\arctg(\frac {\eta\sin(\varepsilon t)}{\varepsilon cosh(\eta x)})
   \end{align}
   and the motion is transformed into
\begin{align}
u=4\arctg(\frac {\eta\sin(\varepsilon( t-vx)/\sqrt{1-v^2})}{\varepsilon cosh(\eta (x-vt)/\sqrt{1-v^2})})
   \end{align}
Thus, in motion there arises a spatial modulation of the form  $\varepsilon vx/\sqrt{1-v^2}$ , similar to de Broglie wave modulation $mvx/\sqrt{1-v^2}$ .
In order soliton-breather particle models adequately describe the particles behavior, they should describe their wave properties manifestation. For example, for the manifestation of the pulson-particles at wave properties interference, the size of such pulson sollutions should be large enough in comparison with microscopic, atomic scale. On the other hand, elementary particles should have sizes of atomic or subatomic scales. This contradiction can be overcome using the assumption of pulson solutions variable size. In the presence of a strong external field, they can be of a small scale, and in free state – the scale determined by the possibility of interference observation. Such pulson solution behavior can be obtained by introducing in the equation of the form (5) an external field
\begin{align}
u_{xx}+u_{yy}+u_{zz}-u_{tt}=\beta V(u)+\alpha u^{\frac{m}{n}}(u_{x}^2+u_{y}^2+u_{z}^2-u_{t}^2)^{\frac{k}{s}}
 ,\label{eq:1}
   \end{align}
Here the coefficient at the nonlinear part $\alpha$  must be sufficiently small compared to the coefficient $\beta$ . When $\alpha <<\beta$  we get Klein-Gordon wave equation with a potential field that is used as a relativistic analog of the Schrodinger equation for a scalar field.

\begin{align}
 i \Psi_{t}=-\frac{\bigtriangleup \Psi}{2m}+V\Psi                      \label{eq:1}
   \end{align}

where $\hbar=1$.  In free state the size of the localized solution will be determined by the equation
\begin{align}
u_{xx}+u_{yy}+u_{zz}-u_{tt}=\alpha u^{\frac{m}{n}}(u_{x}^2+u_{y}^2+u_{z}^2-u_{t}^2)^{\frac{k}{s}}
 ,\label{eq:1}
   \end{align}
Equation (7) is selected as an illustration of such a model (4) and makes no pretense to full and complete description. Authors fully take into account the problem of solutions exponential growth at the external field value of  $V<0$.

One of the consequences of such a model will be a potential barrier width limitation at breathers tunneling. In this case, the breathers tunneling is understood as moving breather transmission through the region of the potential barrier that excesses capacity boosting the breather. Note that the process of potential barrier overcoming by particle is a problem, which is of great importance in many areas of physics [17]. In the case of the classical process the transmission process is possible only when there is an external disturbance. Quantum particle can perform tunneling process with a cetain final probability, generally greater than zero but less than unity. However, what will happen if we take for the consideration of the transmission through the barrier an extended object, for instance, a kink or a breather, instead of a point particle? The answer to this question is important, for example, for fluxons motion in long Josephson contacts with impurities, for the domain walls dynamics of ferroelectric and magnetic materials, as well as many other physical systems in which solitons move in the potential created by irregularities and outside forces [18-21]. It is known that solitons can behave as classical particles in some physical systems [20], but consideration of a soliton as an extended object leads to more interesting phenomena. Among these phenomena we distinguish soliton tunneling, previously studied for such Klein-Gordon equations as the phi four and sine-Gordon equation [21,22].

Next, let us consider the dynamics of breather transmission through the potential barrier region in more detail. Although the equation (4) has a three-dimensional localized solutions, for numerical calculations it is more convenient to consider a one-dimensional equation of the form
\begin{align}
u_{xx}-u_{tt}=V(x)u+ u^{\frac{3}{13}}(u_{x}^2-u_{t}^2)^{\frac{2}{13}}
 ,\label{eq:1}
   \end{align}

In the numerical study there was used an ordinary pseudospectral method, previously applied for the Klein-Gordon equation solution [13,16]. The solution was decomposed into a series of trigonometric functions. Then, the expansion coefficients were found by a time-differential method using the Runge-Kutta method of the fourth order. The number of functions was about 600-900 for the finite length $R = 20-30$. Let us study the solution of equation (10) for the potential field of the form $V=V_0(1-\lambda x)$  where $x<l_0$  and $V=V_1$  while $l_1<x<l_2$  .  Field of the form $V=V_0(1-\lambda x)$  will accelerate the localized solution, and the field  $V_1$ will be a potential barrier.
\begin{figure}[h!]
\center
\includegraphics[width=8 cm, height=5cm]{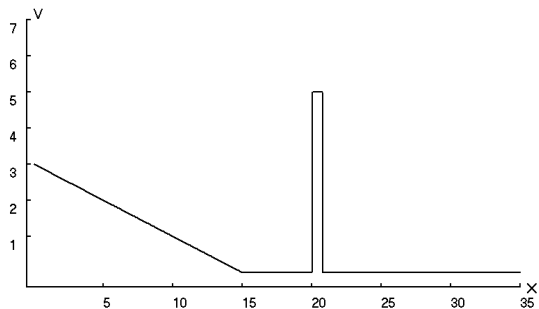}
\caption{ Potential energy, as a function of x with  $V_0=3;\lambda=0.2/3;V_1=5;l_0=15;l_1=20;l_2=20.5$}
\label{schema}
\end{figure}

In the analyzed examples (Fig.2-Fig. 4) for the given parameters - $V_0=3;\lambda=0.2/3;V_1=5;l_0=15$ potential barrier width varied – 1-$l_0=20;l_1=20.5$  ; 2 -$l_0=20;l_1=20.75$  ;3 -$l_0=20;l_1=21$  . Fig. 1b shows the potential considered in our case, i.e. the cases when the potential barrier was higher than the accelerated breather potential.

\begin{figure}[h!]
\center
\includegraphics[width=8 cm, height=5cm]{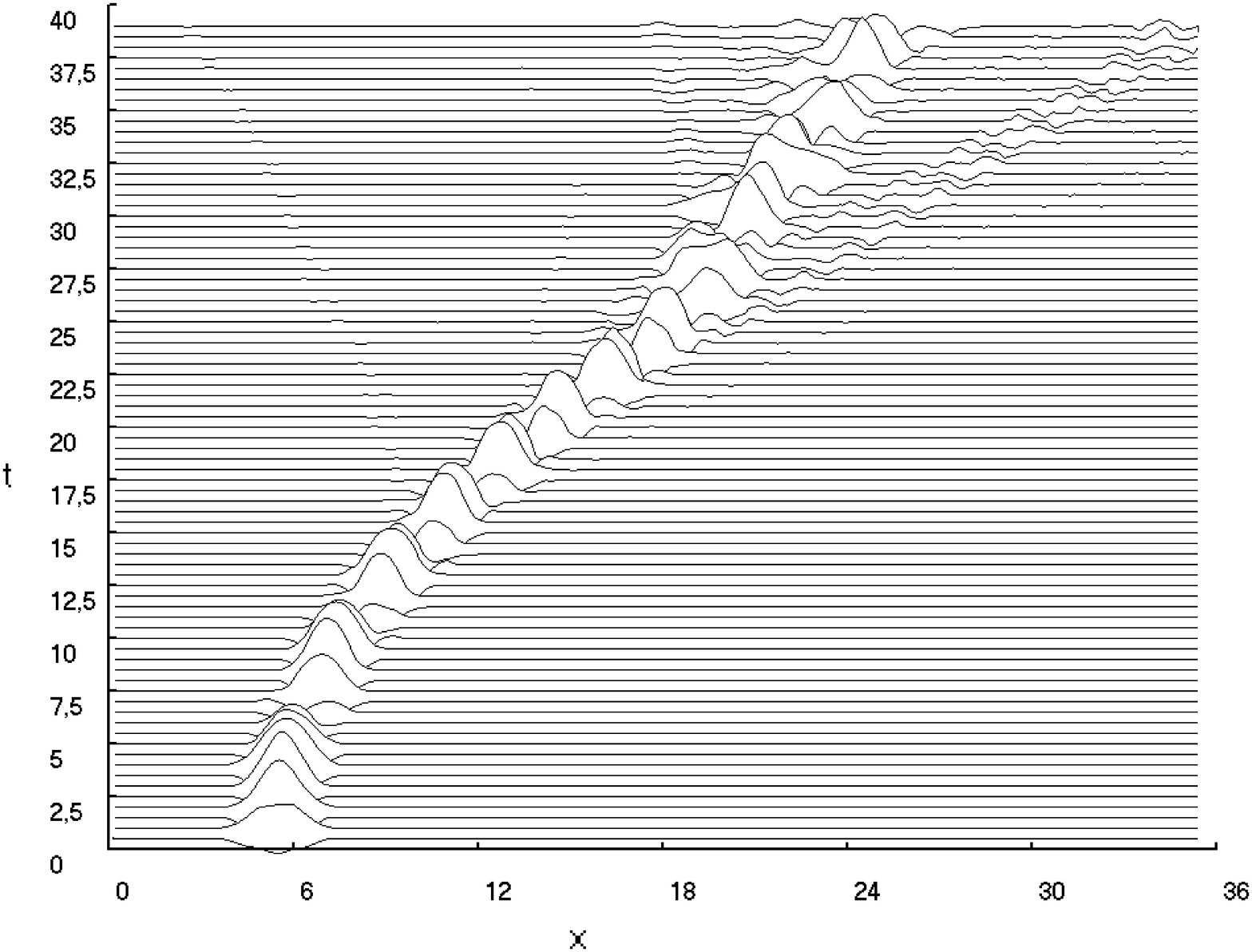}
\caption{ Breather transmission through the barrier at $V_0=3;\lambda=0.2/3;V_1=5;l_0=15;l_1=20;l_2=20.5$ }
\label{schema}
\end{figure}

\begin{figure}[h!]
\center
\includegraphics[width=8 cm, height=5cm]{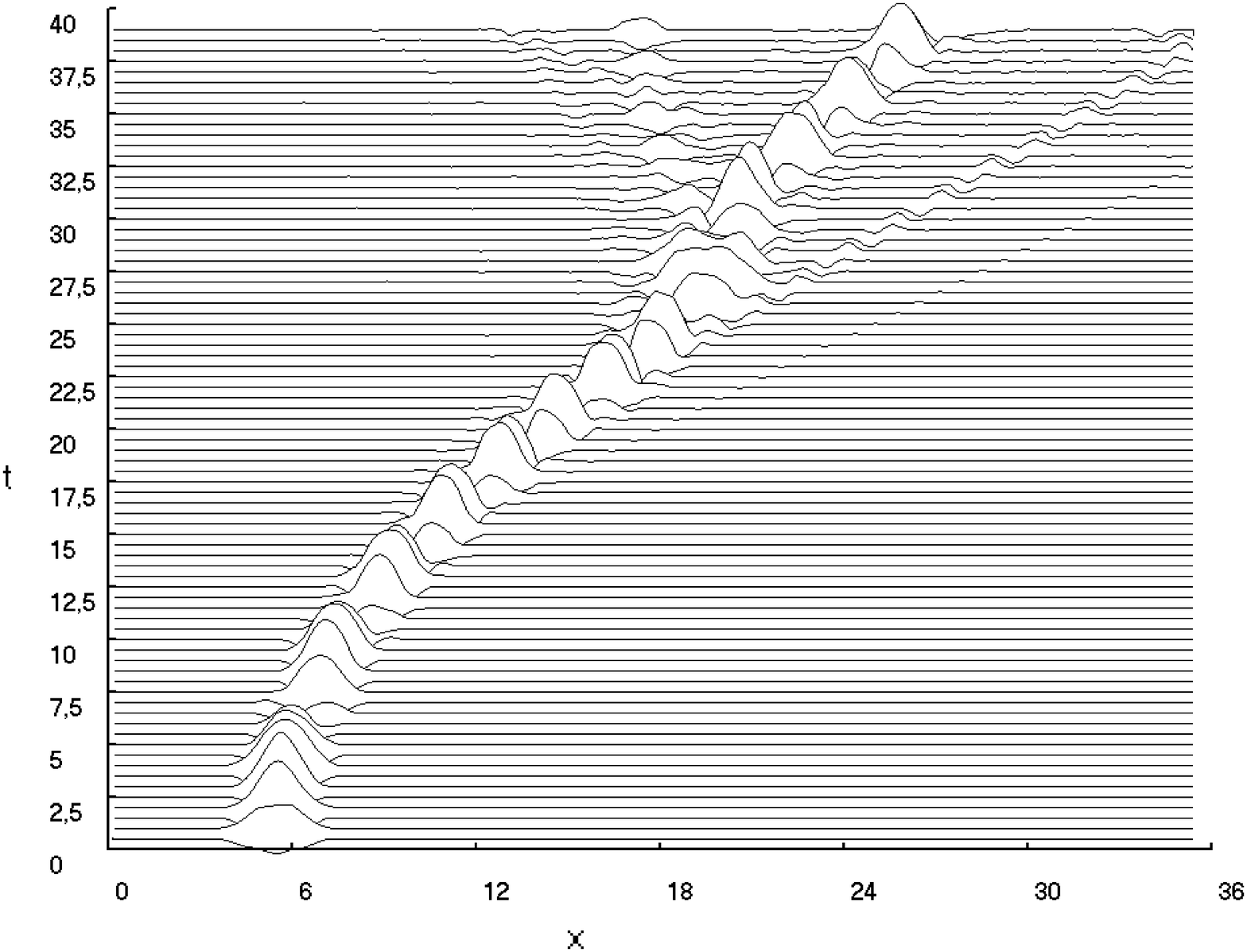}
\caption{ Breather transmission through the barrier at $V_0=3;\lambda=0.2/3;V_1=5;l_0=15;l_1=20;l_2=20.75$ }
\label{schema}
\end{figure}

\begin{figure}[h!]
\center
\includegraphics[width=8 cm, height=5cm]{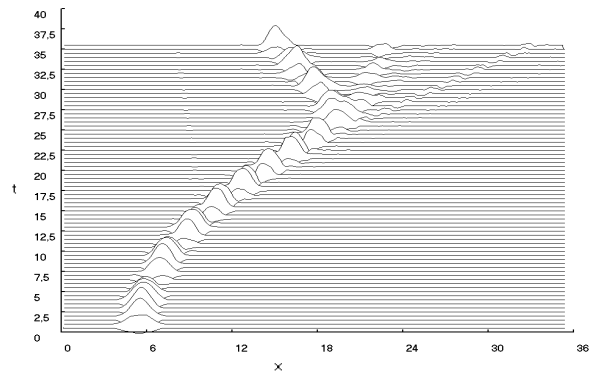}
\caption{ Reflection of a breather from the barrier at $V_0=3;\lambda=0.2/3;V_1=5;l_0=15;l_1=20;l_2=21$  }
\label{schema}
\end{figure}

It is clear from the figures that at motion with constant acceleration (up to the view disturbing barrier solution) the breather does not radiate. Thus, the breather behaves as a material particle in a potential field. Therefore, we can assume that a rectangular barrier will be a forbidden region for breather. However, the figures also show that the breather either passes through the barrier almost completely (i.e., there is the process of its tunneling), or is reflected. Obviously, in this case the passage or reflection from the barrier depends on the barrier width. When the barrier width is greater than a certain critical value, associated with the breather spatial dimensions, it is reflected. During the collision of the breather with a potential barrier there observed radiation. At the same time a decrease in the breather speed is seen.

Although in this one-dimensional case (see Fig.1a) part of the solution may even split from the breather, in a space three-dimensional case, such localized solution splitting will be stopped by the field $r^\frac{n-m}{n}v^\frac{m}{n}$ . That is, a three-dimensional case can be expected to have the solution consisting only from a breather and radiation moving at the limit in a given system (or "light") speed.

Thus, the use of (3+1) D equations of the form (4) or any other relativistic equations having solutions with zero and non-zero mass, provides excellent opportunities to build a different from quantum model of particles tunneling. In three-dimension case one can expect that the solution will consist only of an undivided breather solution and radiation, moving at a "light" speed. Therefore, in the three-dimensional case of this model the localized solution can be interpreted not as a wave amplitude of probability of finding a point particle, but as the particle itself of a finite size. When the scale of the potential field change is bigger than the breather spatial size, the breather behaves as a material particle. Breather exhibits wave properties only when the scale of the external field changes is smaller than its size. In this case the radiation produced at the collision of the breather with a potential barrier, can be interpreted as a non-soliton part of the solution. For nonlinear models that do not have "light" decisions moving at the top speed in a system equal unit in our case, this model should not be allowed. Indeed, for example, in numerical consideration of the nonlinear Schrodinger equation localized solution reflection from the barrier, part of the localized solution will penetrate through the barrier. In this case only probability interpretation of such behavior is possible.

Other, less "radical" particles soliton model requires certain nonlinearity in Schrodinger equations, leading to solutions localization. For example, in the one-dimensional case it is a nonlinear Schrodinger equation

\begin{align}
 i \Psi_{t}=-\frac{\bigtriangleup \Psi}{2m}-\Psi|\Psi|^2+V\Psi                      \label{eq:1}
   \end{align}

Or in the three-dimensional case, a non-linear equation with the fractional degree potential

\begin{align}
 i \Psi_{t}=-\frac{\bigtriangleup \Psi}{2m}+\frac{\Psi}{|\Psi|^\frac{m}{n}}+V\Psi                      \label{eq:1}
   \end{align}
where $n>m;n=2k+1;k=1,2,3,...;m=2l;l=0,1,2,3,...$.

Equation (13) leads to solutions localization in a three-dimensional spherically symmetric case. Numerical study of the equation solutions (13) with potential $V=0$  shows the presence of long-living, with no tendency to spreading, spherical pulson solutions for values $m=2;n=3$ . In this case, the wave function can be regarded as a probability wave amplitude. The differences from tunneling, described by a common Schrodinger equation will be shown, for example, in the following case. Let us consider potential energy $V(r)$  in the following form $V=0$  and when $r<r_1;r>r_2$ , $V=V_0$ , $r>=r_1;r<=r_2$  and  . We place in the area  , bounded by the barrier, a localized solution and numerically trace its evolution, described by equation (13) for the case of spherical symmetry.

\begin{align}
 i \Psi_{t}=-\frac{ \Psi_{rr}}{2m}-\frac{ \Psi_{r}}{mr}+\frac{\Psi}{|\Psi|^\frac{m}{n}}+V\Psi                      \label{eq:1}
   \end{align}

If the localized solution dimensions are smaller than the area before the barrier, the localized solution will remain inside unchanged indefinitely. If the region size is smaller than the localized solution size, localized solution tunneling outside the barrier limits will be thus possible.

In conclusion, we note that the non-linearity inclusion in the consideration of the particles tunneling process leads to some fundamental differences from the linear case:
1) termination of tunneling under the barrier width greater than a certain critical value.
2) decrease in the breather speed after the potential barrier passing and appearance of radiation
3) a sharp breather localization in a potential well with the size of the potential well greater than some critical value.
The first difference is connected with the fact that at the linear quantum-mechanical description the particle tunneling probability will decrease exponentially with the barrier width, and will not stop abruptly. That is, in the ordinary quantum-mechanical description there will not be any drastic tunneling changes with increasing width of the potential well. The second difference is due to the fact that in the quantum-mechanical description of the tunneling transition through the potential barrier, the reflected wave has the same energy as the incident and there is no radiation in this case. Thus, if the breather model of particles description is applicable, there must appear the abovementioned imperative corollary - 1), 2) and 3).
If we assume, for example, that the soliton-breather particles model is applied for electrons, these corollaries will have to manifest themselves at tunneling electron emission from metal. Corollary 1) in this case would lead to a discontinuous change in the tunneling current value at reaching certain critical values of the potential barrier width. Corollary 2) would lead to the appearance of radiation at tunnel emission. Corollary 3) would also lead to a discontinuous change in the tunneling current at reaching sample critical dimensions, from which the emission occurs. Although the electron emission is a well-studied area, such experiments have not been specially carried out, and the authors do not have any information about monitoring of such effects. Note that the soliton model of particles tunneling described by equations (5) and (9) has a methodological interest and requires further study. We would also want to underline that the imperative corollaries from the soliton-breather particles models, associated with equations of the form (4) and (9), describe new features, that don’t appear when using the linear quantum-mechanical description. It can be stated that these corollaries are fairly common for breather-soliton particles model, and their verification could help to answer the question of such models applicability in real physical systems.



\begin{thebibliography}{99}





\bibitem{bib:asc}
A.\,Scott  (ed.). Encyclopedia of Nonlinear Science,New York: Routledge, (2004).

 \bibitem{bib:br}
 O.\,Braun, Y.\, Kivshar Y.The Frenkel-Kontorova Model: Concepts, Methods, and Applications
. Springer:Verlag Berlin Heidelberg, 2004

 \bibitem{bib:td}
T.\,Dauxois ,M.\, Peyrard, Physics of solitons. New York: Cambridge University Press, 2010.

 \bibitem{bib:cjk}
C.\, J.\, K.\,Knight, G.\, Derks,A.\, Doelman , H.\, Susanto ,Journal of Differential Equations {\bf 254}, 408–468,(2013).
\bibitem{bib:fp}

  J.\,Cuevas-Maraver,P.\, Kevrekidis,F.\, Williams, The Sine-Gordon Model and Its Applications: From Pendula and Josephson Junctions to Gravity and High-energy Physics, vol. 10. Springer, 2014.


\bibitem{bib:ds}
D.\, Saadatmand,J.\, Kurosh, Braz J Phys {\bf 56},43,(2013).

\bibitem{bib:jag}
J.\,A.\, Gonzalez, A.\, Bellorin, L.\,E.\, Guerrero, Solitons and Fractals {\bf 33}, 143, (2007).


\bibitem{bib:alf}
G.\,L.\, Alfimov, W.\,A.\,B.\,Evans, L.\,Vazquez, Nonlinearity {\bf 13}, 1657, (2000).

\bibitem{bib:pav}
S.\,A.\ Pavliuchenko, A.\,V.\ Toporensky, Gravitation  \& Cosmology, {\bf 6}, 241, (2000).

\bibitem{bib:komd}
V.\, Koutvitsky, E.\, Maslov, Phys. Rev. D, {\bf 83}, 124028,(2011).

\bibitem{bib:mach}
V.\, Koutvitsky, E.\, Maslov, Journal of mathematical physics, {\bf 47}, 022302, (2006).

\bibitem{bib:scr}
T.\,H.\, Skyrme ,Proc. Roy Soc. London.  Vol. {\bf A247}, 260–278,(1958).




\bibitem{bib:rd}
R.\,Dodd , J.\, Eilbeck, J.\, Gibbon and H.\, Morris ,  Solitons and nonlinear wave equations,  Academic Pr (1984).




\bibitem{bib:vgm}
V.\, G.\, Mahankov, Particles and Nuclei, {\bf 14}, 123-180, (1983).

\bibitem{bib:eksa}
E.\,G.\,Ekomasov, R.\,K.\,Salimov, Jetp letters, {\bf 100}, 7,477-480, (2014)


\bibitem{bib:eks}
E.\,G.\,Ekomasov, R.\,K.\,Salimov, Jetp letters, {\bf 102}, 2,122, (2015)


\bibitem{bib:ht}
P.\, Hanggi, P.\, Talkner, M.\, Borkovec, Rev. Mod. Phys.{\bf 62}, 251,(1990).

\bibitem{bib:dp}
T.\, Dauxois ,M.\, Peyrard. Physics of solitons. New York: Cambridge University Press, 2010.

\bibitem{bib:ckw}
J.\, Cuevas-Maraver, P.\, G.\, Kevrekidis, F.\, Williams, Editors. The sine-Gordon model and its applications. From pendula and josephson junctions to gravity and high-energy physics. Springer, Heidelberg, New York, Dordreht, London, 2014.
\bibitem{bib:klb}
G.\, Kalbermann, Phys. Rev. E {\bf 55}, R6360, (1997).
\bibitem
J.\, A.\, Gonzales, A.\, Bellorin,  L.\, E.\, Guerrero. Phys Rev E {\bf 60},1, R37, (1999).







\end{thebibliography}
\end{document}